\documentclass[10pt,journal,compsoc]{IEEEtran}
\usepackage{graphicx} 
\usepackage[switch]{lineno}

\ifCLASSOPTIONcompsoc
  \usepackage[nocompress]{cite}
\else
  \usepackage{cite}
\fi
\ifCLASSINFOpdf
\else
\fi
\usepackage{booktabs}
\usepackage{amsmath}
\usepackage{bm}
\usepackage[colorlinks=true, linkcolor=black, citecolor=black, urlcolor=black]{hyperref}
\usepackage{amssymb}
\usepackage{multirow} 
\begin{document}
% \linenumbers
\title{Denoising Pre-Training and Customized Prompt Learning for Efficient Multi-Behavior Sequential Recommendation}
\newcommand{\Email}[1]{\texttt{\href{mailto:#1}{#1}}}
\author{Hao~Wang,
        Yongqiang~Han,
        Kefan~Wang,
        Kai~Cheng,
        \\
        Zhen~Wang,
        Wei~Guo,
        Yong~Liu,
        Defu~Lian,
        and~Enhong~Chen,~\IEEEmembership{Fellow}

\IEEEcompsocitemizethanks{
  \IEEEcompsocthanksitem H.~Wang, Y.~Han, K.~Wang, K.~Cheng, D.~Lian and E.~Chen~(corresponding author) are with the Anhui Province Key Laboratory of Big Data Analysis and Application, School of Computer Science and Technology, University of Science and Technology of China, Hefei, Anhui, 230026, China.
  \protect\Email: \{wanghao3, liandefu,cheneh\}@ustc.edu.cn, \{harley,wangkefan,ck2020\}@mail.ustc.edu.cn.

  \IEEEcompsocthanksitem Z.~Wang is a researcher from SUN YAT-SEN University.
  \protect\Email: \{wangzh665\}@mail.sysu.edu.cn.
  
  \IEEEcompsocthanksitem W.~Guo and Y.~Liu are researchers at Huawei’s Noah’s Ark Laboratory.
  \protect\Email: \{guowei67,liu.yong6\}@huawei.com.
}  

}


\markboth{Journal of \LaTeX\ Class Files,~Vol.~14, No.~8, August~2015}%
{Shell \MakeLowercase{\textit{et al.}}: Bare Demo of IEEEtran.cls for Computer Society Journals}



\IEEEtitleabstractindextext{%
\begin{abstract}
In the realm of recommendation systems, users exhibit a diverse array of behaviors when interacting with items. This phenomenon has spurred research into learning the implicit semantic relationships between these behaviors to enhance recommendation performance. However, these methods often entail high computational complexity. To address concerns regarding efficiency, pre-training presents a viable solution. Its objective is to extract knowledge from extensive pre-training data and fine-tune the model for downstream tasks. Nevertheless, previous pre-training methods have primarily focused on single-behavior data, while multi-behavior data contains significant noise. Additionally, the fully fine-tuning strategy adopted by these methods still imposes a considerable computational burden. In response to this challenge, we propose DPCPL, the first pre-training and prompt-tuning paradigm tailored for Multi-Behavior Sequential Recommendation. Specifically, in the pre-training stage, we commence by proposing a novel Efficient Behavior Miner (EBM) to filter out the noise at multiple time scales, thereby facilitating the comprehension of the contextual semantics of multi-behavior sequences. Subsequently, we propose to tune the pre-trained model in a highly efficient manner with the proposed Customized Prompt Learning (CPL) module, which generates personalized, progressive, and diverse prompts to fully exploit the potential of the pre-trained model effectively. Extensive experiments on three real-world datasets have unequivocally demonstrated that DPCPL not only exhibits high efficiency and effectiveness, requiring minimal parameter adjustments but also surpasses the state-of-the-art performance across a diverse range of downstream tasks.
\end{abstract}

}

\maketitle

\IEEEdisplaynontitleabstractindextext

% \fi
\IEEEpeerreviewmaketitle

% \IEEEraisesectionheading{\section{Introduction}\label{sec:introduction}}

% for IEEE Computer Society journal papers produced under \LaTeX\ using
% IEEEtran.cls version 1.8b and later.
% % You must have at least 2 lines in the paragraph with the drop letter
% % (should never be an issue)
% I wish you the best of success.

\section{INTRODUCTION}
% With the rapid advancement of the Internet, recommendation systems have become ubiquitous on online platforms. Among these systems, sequential recommendation (SR) has garnered widespread attention from both academic researchers and industry practitioners. SR involves predicting the next item for users by treating their historical interactions as temporally-ordered sequences. This approach has been extensively studied in various works such as \cite{seqrec1, TiSASRec, LESSER, seqrec2, CBiT, MCLSR, GUESR, SRGNN}. In actuality, users exhibit a multitude of behaviors when interacting with items, which reflect their multifaceted preferences. For example, on e-commerce platforms, users can engage with items through diverse behaviors such as clicking, tagging as favorites, adding to carts, and making purchases. These varied behaviors represent users' preferences across multiple facets, and have been utilized as supplementary knowledge to enhance the accuracy of recommendations for the target behavior. This approach has been explored in studies such as \cite{multibehavior1, MMCLR, TGT, KMCLR}.

\IEEEPARstart{W}{ith} the rapid evolution of the Internet, recommendation systems have proliferated across online platforms. Among these systems, sequential recommendation (SR) has garnered considerable attention from both academic researchers and industry practitioners. SR involves predicting the next item for users by analyzing their historical interactions as temporally-ordered sequences. This approach has been extensively explored in various studies such as \cite{q1, q3, seqrec1, TiSASRec, LESSER, seqrec2, CBiT, MCLSR, GUESR, rec_tkde}.

In reality, users exhibit a diverse range of behaviors when interacting with items, reflecting their multifaceted preferences. For instance, on e-commerce platforms, users may engage with items through various behaviors like clicking, tagging as favorites, adding to carts, and making purchases. These diverse behaviors encapsulate users' preferences across multiple dimensions and have been leveraged as supplementary knowledge to enhance recommendation accuracy for the target behavior. This approach has been investigated in studies such as \cite{multibehavior1, MMCLR, TGT, KMCLR}.

\begin{figure}
  \includegraphics[width=0.48\textwidth]{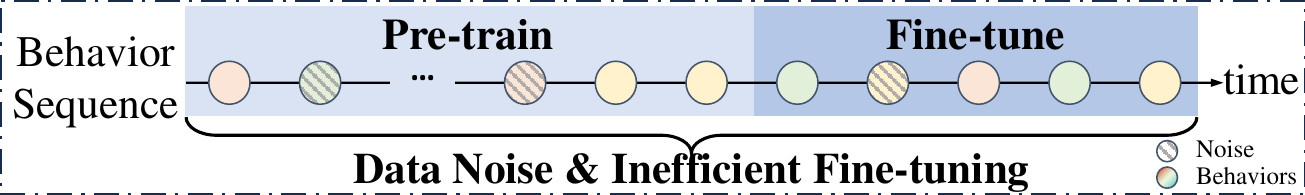}
    \caption{Illustration of our motivations: the rapid increase in user interactions leads to a substantial amount of data, which introduces significant noise. This noise presents challenges in accurately capturing user preferences and highlights the need for more robust pre-training and fine-tuning methods for efficient transfer to downstream tasks.}
    \label{FIG_motivation}
\end{figure}

% Despite the considerable achievements made in recent years, the deployment of these methods is severely limited by online latency due to their high computational complexity. This limitation results in a compromise on model input length and inferior results. In order to address this issue, some pre-training methods such as PeterRec~\cite{PeterRec}, UPRec~\cite{UPRec}, and PinnerFormer~\cite{PinnerFormer} have been proposed. These methods employ large architectures such as transformers to learn common knowledge from large historical data sets through self-supervised tasks, and then fine-tune the model for different downstream tasks~\cite{PLCR, CPTPP}.

Despite the considerable achievements made in recent years, the deployment of these advanced methods is severely constrained by online latency due to their high computational complexity. This limitation forces a compromise on the model input length, leading to suboptimal results. To address this issue, several pre-training methods have been proposed, including PeterRec~\cite{PeterRec}, UPRec~\cite{UPRec}, and PinnerFormer~\cite{PinnerFormer}. These methods leverage large architectures, such as transformers, to learn common knowledge from extensive historical datasets through self-supervised tasks. After pre-training, the model is fine-tuned for different downstream tasks using recent user behavior data, as discussed in studies such as~\cite{PLCR, CPTPP, PinnerFormer}. This approach allows models to effectively utilize vast amounts of historical data while maintaining efficiency and improving performance on specific tasks.

With the advent of the pre-training paradigm, the issue of deployment complexity has been alleviated to some extent. This paradigm has enabled models to handle extensive historical data more efficiently, thus improving performance across various downstream tasks. However, despite achieving significant success, as illustrated in Figure~\ref{FIG_motivation}, these methods still overlook certain problems that need to be addressed, such as noise in historical behaviors and the tuning efficiency and effectiveness in recommendation scenarios.

\textbf{\textit{Noise in historical behaviors.}} In multi-behavior sequential recommendation scenarios, the data usually covers a variety of user behaviors such as clicking, browsing, and purchasing, which are interspersed with a large amount of noise information, especially the clicking behavior. These noises can negatively affect the pre-trained model, making it difficult to accurately capture the user's behavioral patterns, or even overfitting to the noisy information, which ultimately results in unsatisfactory downstream performance. Towards this problem, some pre-training methods \cite{S3Rec, CL4SRec} proposed to enhance the robustness of the model by contrastive learning with different data augmentation strategies. However, these methods will introduce extra noise during the random data augmentation process, and potential misalignment between the self-supervised task and the recommendation task further hinders their application in the high-noise scenarios.

\textbf{\textit{Tuning Efficiency and Effectiveness In recommendation scenarios.}}
Although promising results are achieved in downstream tasks, the comprehensive fine-tuning strategy adopted by previous methods usually requires more time and computational resources. In addition, pre-trained models can learn rich user item representations on large-scale data, but full fine-tuning may cause the catastrophic forgetting issue \cite{catastrophic_forgetting, cf1}, i.e., some common knowledge acquired from pre-training is forgotten. Therefore, instead of full fine-tuning, we propose to adopt the prompt-tuning technique \cite{GPT3} to utilize the pre-trained model, which is proven to be efficient and effective. However, it is non-trivial to adapt prompt-tuning to the multi-behavior sequential recommendation scenario. Unlike NLP, the tokens of the pre-trained recommendation model lack semantic information, making it challenging to manually design hard prompts. Besides, general prompts fail to fulfill the personalization requirement in recommendation systems.

To tackle the two aforementioned challenges, we propose a novel \textbf{D}enoising \textbf{P}re-training and \textbf{C}ustomized \textbf{P}rompt \textbf{L}earning paradigm (DPCPL) for the efficient multi-behavior sequential recommendation. \textbf{Firstly}, to alleviate the noise problem in user behavior history, we innovatively propose an Efficient Behavior Miner (EBM) in the pre-training stage, which can decompose the user behavior at different time scales through Fast Fourier Transform~\cite{FFT}. In addition, we adopt learning filter kernel and techniques such as frequency-aware fusion and chunked diagonal mechanism to reduce noise at different scales and minimize the number of parameters in the model. \textbf{Secondly}, for efficient tuning of pre-trained models in multi-behavioral scenarios, we point out the necessity of personalization for prompts in recommendation systems and introduce a new concept, Customized Prompt Learning (CPL), which dynamically sets up the prompts based on a variety of user-specific auxiliary information. Furthermore, we find the pre-training model adopts a bottom-up process to gradually condense user interests from user behaviors, for this reason, we propose the Prompt Factor Gate (PFG) structure to leverage users' personalized information to generate layer-wise prompts, enabling an in-depth impact on the model's learning process. Besides, to avoid homogeneity of prompts, we introduce a compactness regular loss function to better control the diversity of prompts. The contributions of this paper can be summarized as follows:

(1).  We present a comprehensive study on the challenges of pre-training and prompt-tuning in multi-behavior sequential recommendation scenarios. Our study highlights the crucial importance of denoising historical behaviors and implementing personalized prompt learning to enhance the accuracy and efficiency of recommendations.

(2).  We propose an efficient denoising module capable of effectively denoising information across multiple scales of behavior sequences with frequency domain mapping during the pre-training stage.

(3).  We introduce the concept of customized prompt learning, aimed at generating personalized, progressive, and diverse prompts to fully utilize the potential of pre-trained models in multiple behaviors.

(4).  We conducted extensive experiments on three real-world datasets and have demonstrated the effectiveness and efficiency of DPCPL, which consistently achieves superior performance across various downstream tasks with minimal parameter tuning.

\section{RELATED WORK}
\subsection{Multi-behavior Sequential Recommendation}

Recommendation system recommends personalized content based on individual preferences, sequential recommendation predicts a user's next target item based on their historical behavior, playing a crucial role in enhancing the user experience on online platforms \cite{q4, q5, q6, q7, q8, q9}. With the emergence of deep learning, sequential recommendation models such as BERT4Rec \cite{BERT4Rec}, DIN \cite{DIN}, SASRec \cite{SASRec}, and FEARec \cite{FEARec} were introduced for recommendation tasks. However, they failed to consider the diversity of user interactions in real-world scenarios, such as clicking, liking, and purchasing in e-commerce, which provided valuable insights into user intent. To overcome this limitation, researchers have proposed various methods for handling multi-behavior data.

Previous research has explored the use of multi-task frameworks to optimize recommendation systems. One approach is to model the cascade relationship among different user behaviors, as done in NMTR \cite{NMTR}. Another approach is to assign user behaviors to distinct tasks and employ hierarchical attention mechanisms to improve recommendation efficiency, as in DIPN \cite{DIPN}. Other studies have focused on enhancing recommendation by fusing multi-behavior data and using other behaviors as auxiliary signals. This has been achieved through attention mechanisms \cite{atten-rec1}, graph neural networks \cite{graph-rec1, MBGMN, atten-rec2}, or other related approaches. For example, MATN \cite{MATN} used a transformer and gated network to capture behavior relationships, while CML \cite{CML} introduced a multi-behavior contrastive learning framework to enhance behavior representations. KMCLR \cite{KMCLR} utilized comparative learning tasks and functional modules to improve recommendation performance through the integration of multiple user behavior signals.

Although the fusion of multi-behavior information can further effectively explore user behavior patterns and multi-dimensional user interests, some behavior data will inevitably introduce noise to the modeling of user interests. This noise poses challenges to the judgment of user sequence interests. Moreover, with the diversification of behavior data, user behavior sequences become increasingly lengthy in a short period, which presents challenges to the efficiency of the sequence recommendation model.

\subsection{Pre-trained Recommendation Methods}
Nonetheless, existing multi-behavior recommendation methods face significant challenges. Firstly, in-depth exploration of user interests often leads to increased model complexity, which compromises operational efficiency in real-world applications. Secondly, the proliferation of behaviors extends the sequences of user interaction history, complicating effective modeling. A practical solution to these problems is to use pre-trained models, which can avoid heavy computational demands during actual deployment.

Therefore, some pre-training methods have been proposed to address computational complexity problem~\cite{DUPN, IERT, ASREP}, which employ large architectures such as transformers to learn common knowledge from large historical data sets through self-supervised tasks, and then fine-tune the model for different downstream tasks to alleviate the issue of deployment complexity. To exemplify, $S^3$-Rec \cite{S3Rec} employs pre-training to bolster data augmentation, with a focus on deciphering correlations across attributes, items, subsequences, and sequences through the lens of mutual information maximization principles. PeterRec \cite{PeterRec} starts by pre-training a comprehensive model and then integrates specialized sub-models for downstream tasks, allowing for easy task adaptation. UPRec \cite{UPRec} uses user attributes and social graphs in pre-training to enhance personalization, while CL4SRec \cite{CL4SRec} employs contrastive learning for sequence modeling. PinnerFormer \cite{PinnerFormer} focuses on long-term action prediction with a dense all-action loss. 

However, the self-supervised tasks designed by some methods may introduce additional noise during the random data augmentation process. Additionally, the potential misalignment between self-supervised tasks and recommendation tasks further limits their effectiveness in high-noise scenarios. Moreover, these methods often overlook the need for efficient fine-tuning of the pre-trained models to facilitate the effective transfer to downstream tasks.

\subsection{Prompt Learning in Recommendation}
% In contrast to earlier fine-tuning methods, the success of GPT \cite{GPT3} has underscored the viability of maintaining a well-trained model's integrity while modifying solely the input prompt. In recent years, more and more work has begun to apply the prompt-tuning paradigm to the field of sequential recommendation, through the fewer parameters and efficiency of prompt-tuning to improve the model effect. PPR~\cite{PPR} pre-trains recommendation models and uses sequence prompts to enhance model performance. UP5 \cite{UP5} applies fairness-aware prompt templates to improve recommendation equity. PLCR \cite{PLCR} infuses domain knowledge into pre-trained models and fine-tunes with prompts for more effective cross-domain recommendations. DPT \cite{DPT} adopts a three-stage denoising and prompt fine-tuning approach to reduce noise impacts, while CPTPP \cite{CPTPP} utilizes user information and graph-based contrastive learning to align pre-trained user vectors with downstream tasks. However, these methods lack personalized modeling under complex behaviors, resulting in suboptimal performance on multi-behavior tasks.

In contrast to earlier fine-tuning methods, the success of GPT \cite{GPT3} has highlighted the effectiveness of maintaining a well-trained model's integrity while modifying only the input prompt. This has paved the way for the prompt-tuning paradigm~\cite{prompttkde}, which has gained traction in recent years for sequential recommendation tasks. By leveraging fewer parameters and improving efficiency, prompt-tuning enhances model performance.

Several notable works have applied this paradigm. PPR~\cite{PPR} pre-trains recommendation models and employs sequence prompts to boost performance. UP5 \cite{UP5} integrates fairness-aware prompt templates to promote recommendation equity. PLCR \cite{PLCR} incorporates domain knowledge into pre-trained models and fine-tunes them with prompts for more effective cross-domain recommendations. DPT \cite{DPT} utilizes a three-stage denoising and prompt fine-tuning approach to mitigate noise impacts. CPTPP \cite{CPTPP} combines user information with graph-based contrastive learning to align pre-trained user vectors with downstream tasks.

Despite these advancements, these methods often fall short in personalized modeling under complex behavioral scenarios, leading to suboptimal performance in multi-behavior tasks. The lack of nuanced personalization and adaptation to individual user behaviors remains a significant challenge, suggesting the need for further refinement and innovation in this area.

\section{PROBLEM DEFINITION}
In this section, we will first delve into the task formulation of pre-training and fine-tuning for multi-behavioral sequential recommendation. This process involves a two-step approach where we initially pre-train the model on a large dataset to capture general patterns, followed by fine-tuning it on a specific dataset to tailor it to the user's needs. The goal is to provide a more personalized recommendation system that takes into account various behavioral sequences. Following this, we introduce the innovative concept of Customized Prompt Learning. This novel idea aims to leverage the full potential of pre-trained large-scale models by generating prompts dynamically based on individual user characteristics and preferences. By doing so, we can enhance the model's ability to understand and respond to users' unique behaviors, ultimately improving the overall recommendation system's performance.

\begin{figure}
  \includegraphics[width=0.48\textwidth]{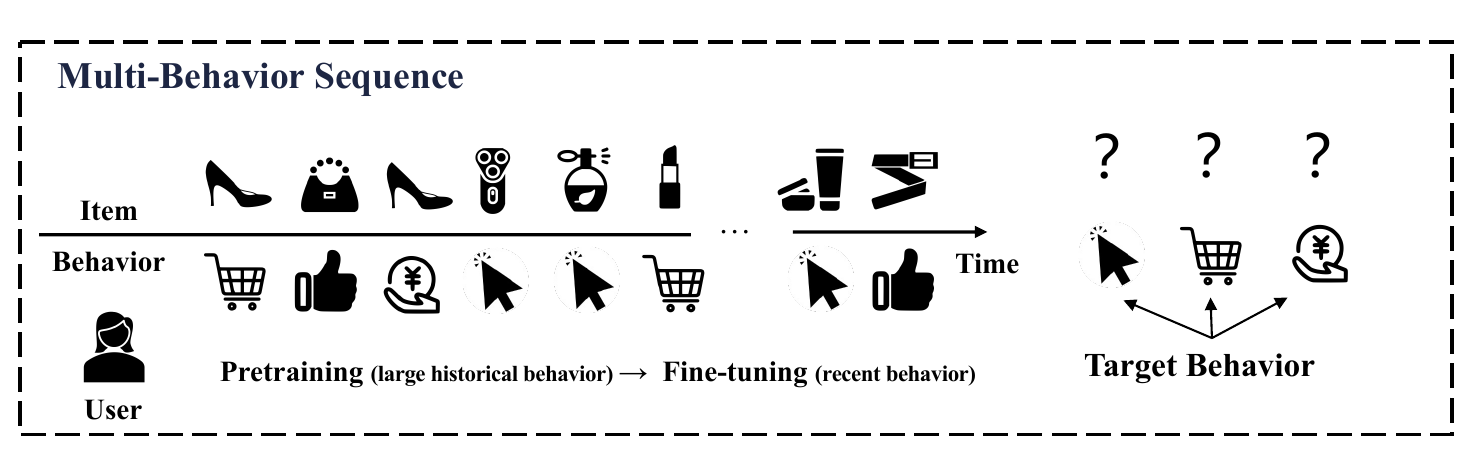}
    \caption{Illustration of our problem: Our task involves taking a sequence of user behaviors as input and predicting the next item of user interaction under a specific target behavior. This target behavior can vary, serving as different downstream tasks.}
    \label{FIG_task}
\end{figure}

In traditional sequential recommendation systems, Multi-BehaviorIn traditional sequential recommendation systems, Multi-BehaviorRecommendation (MBSR) utilizes various forms of user interaction data with items, such as clicking, adding to cart, and favoriting, to capture more nuanced patterns of user behavior. Typically, in a multitude of behaviors, there is only one designated target behavior, such as purchasing, and the predictive task involves forecasting items under this particular behavior. However, due to limitations in online latency, it is often essential to pre-train a model on historical data and subsequently fine-tune it using more recent data in real-world scenarios. The primary objective of MBSR is to enhance the precision of recommendations while simultaneously minimizing latency by effectively utilizing information about user interactions with items. This problem can be formally defined as follows:

    DEFINITION 1. \textit{\textbf{(Pre-training and Fine-tuning for Multi-Behavior Sequential Recommendation)} Given the sets of users \(U\), items \(V\), and types of behavior \(B\), for a user u (\(u \in U\)), his/her behavior-aware interaction sequence \(S_u\) consists of individual triples \((v, t, b)\) which are ordered by time t. Each triple represents the interacted item v under the behavior type b at time \(t\). Then a Pre-training and Fine-tuning paradigm consists of two steps: Firstly, given the previous part of users' behavior-aware interaction sequences $S_u = [(v_{1}, t_{1}, b_{1}), (v_{2}, t_{2}, b_{2}),  \dots, (v_{pre}, t_{pre}, b_{pre})]$, where index $pre$ is the truncated sequence length in pre-training, a model is pre-trained with multiple pretext tasks based on the sequences. Subsequently, the pre-trained model is further fine-tuned on the remaining user sequences $S_u = [(v_{pre+1}, t_{pre+1}, b_{pre+1}), \dots,(v_{|S_u|-1}, t_{|S_u|-1}, b_{|S_u|-1})]$ with a next-item prediction task, which aims to predict the item with the target behavior type at the next time step $(v_{|S_u|}, t_{|S_u|}, b_{\textrm{target}})$.}

However, the comprehensive fine-tuning strategy adopted by previous pre-training methods still poses a considerable computational burden, hindering their application in real-world scenarios. So we intend to follow a pre-training \& prompt-tuning paradigm to model user preferences for the sake of efficiency, which is yet non-trivial. On the one hand, the token of the pre-trained recommendation model is the ID of items, which lacks specific semantics, making it challenging to manually design hard prompts based on the grammatical experience as in NLP. On the other hand, personalization plays a crucial role in the field of recommendation systems, which means prompts should vary for each user. To this end, we point out that prompt learning in recommendation scenarios requires personalization and introduce a new concept called Customized Prompt Learning, which involves dynamically setting prompts based on various user-specific auxiliary information. Formally, we define it as:

  DEFINITION 2. \textit{\textbf{(Customized Prompt Learning)} Given a user \(u \in  U\) and his/her information including attribute sets {\(A_u\)} and behavior-aware interaction Sequence {\(S_u\)}, the output is a set of customized prompts \(\{p^u_1, p^u_2,..., p^u_{n}\}\). \( p^u_{n} \) represents the \( n \)-th prompt vector for user \( u \).These prompts are generated by fusing user-specific information to tune the pre-trained model without modifying its parameters.}

Based on the above problem definition, we propose the model DPCPL to address the problem of noise impact and computational burden in the previous method.  

\begin{figure*}[!t]\centering
	\includegraphics[width=1\textwidth]{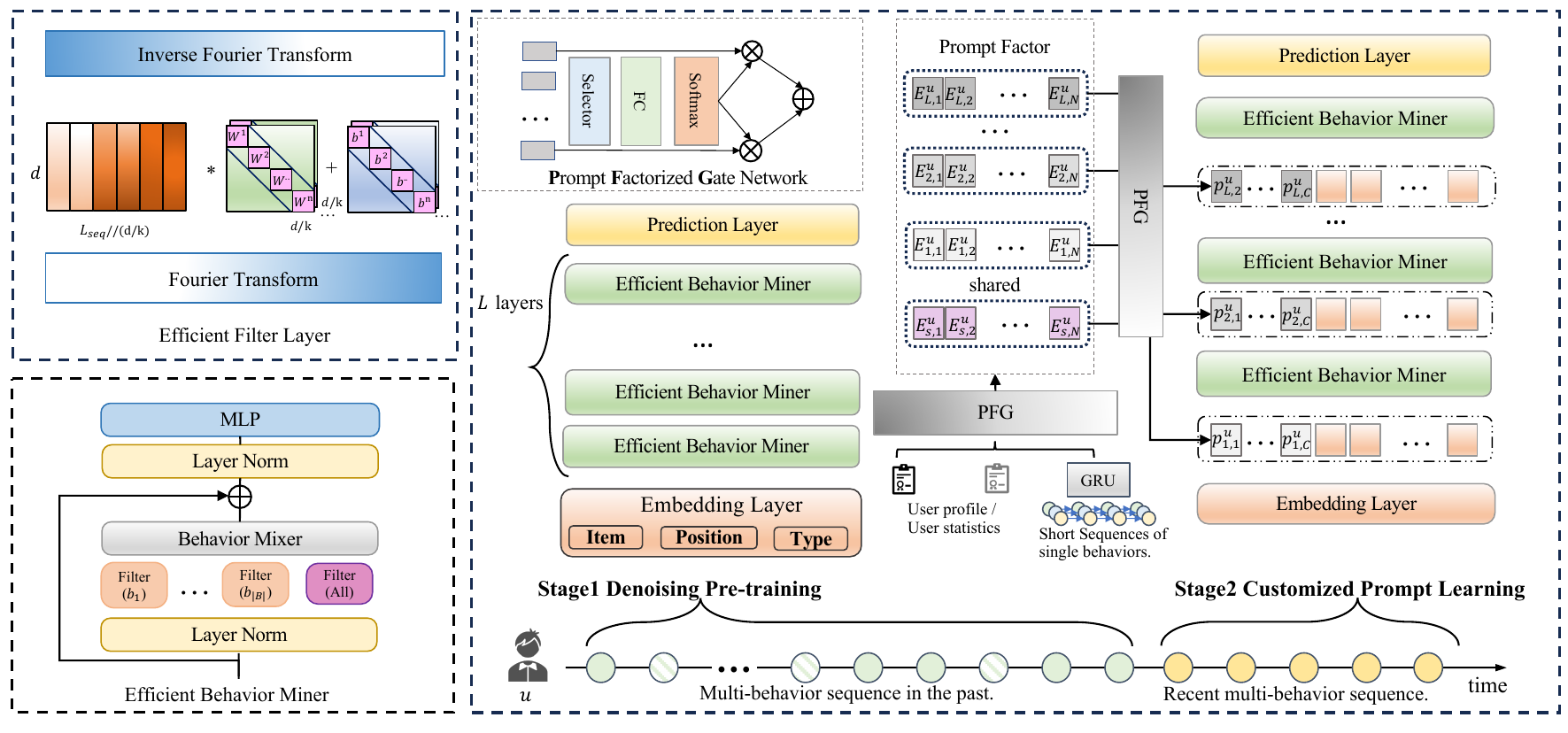}
	\caption{Our DPCPL model, a pre-training and prompt-tuning paradigm tailored for Multi-Behavior Sequential Recommendation. Specifically, in the pre-training stage, we commence by proposing a novel Efficient Behavior Miner (EBM) to filter out the noise at multiple time scales, therefore facilitating the comprehension of the contextual semantics of multi-behavior sequences. Subsequently, we propose to tune the pre-trained model in a highly efficient manner with the proposed Customized Prompt Learning (CPL) module, which generates personalized, progressive, and diverse prompts to fully exploit the potential of the pre-trained model effectively.}
    \label{FIG_main}

\end{figure*}

\section{METHODOLOGY}
In this section, we present our model DPCPL, which mainly consists of a denoised pre-trained module and a prompt learning module with personalization, progressivity, and diversity. Our framework is a pioneering model for solving multi-behavior sequential recommendation problems using customized prompt learning. The overall flow of the model is depicted in Figure~\ref{FIG_main}.

\subsection{Behavior-Aware Sequence Embedding}
The embedding layer of DPCPL integrates item information ($v$), position information ($p$), and behavior information ($b$). For a triad \((v, p, b)\) within a user behavior sequence ($S$), the embedding is denoted as follows:
\begin{equation}
e = e_v + e_p + e_b, \quad S = [e_1,e_2,...,e_{L_{seq}}] \in \mathbb{R}^{L_{seq} \times d},
\end{equation}
where $L_{seq}$ represents the sequence length, and $d$ denotes the embedding size. This embedding combines item information, position information, and behavior information to reflect the user's behavioral sequences more comprehensively, which helps to improve the model's understanding of the user's interests and behaviors, and thus improves the accuracy of personalized recommendations.

\subsection{De-noised Pre-trained Model}
Pre-trained models have been devoted to addressing the long-sequence overload issue \cite{PinnerFormer}. They often pre-train using long-term historical behavior data and then fine-tune using recent behavior data. 
However, in multi-behavior recommendation scenarios, the data typically includes a variety of user behaviors such as clicking, browsing, and purchasing. These behaviors are often interspersed with a significant amount of noise, such as random click activity, which greatly impacts the training of pre-trained models and can further degrade their downstream performance.

Therefore, our research focuses on the need to purify sequences during the pretraining phase. One of the primary challenges in this task is the lack of explicit labels for noise elements, which makes it an unsupervised task. In real-world scenarios, collecting negative feedback from users to identify data noise is extremely difficult. Although previous pre-training methods~\cite{CL4SRec, S3Rec} based on contrastive learning can improve the robustness of models to some extent, they often introduce additional noise during the random data augmentation process. Moreover, the inherent conflict between self-supervised tasks and recommendation tasks further impedes their effectiveness in high-noise environment.

Recently, some research work has revealed the possibility of frequency-domain denoising in recommendation scenarios with a learnable filter \cite{FMLP}. Through extensive comparative experiments, it has been conclusively demonstrated that different frequency components of user sequences have varying impacts on recommendation effectiveness. Specifically, low-frequency components tend to contain valuable information, while high-frequency components are often data noise, which is an established conclusion. At the same time, multiplication in the frequency domain can replace convolution in the time domain, fully integrating various feature information while denoising.

Inspired by the above findings, we transform the user behavior sequence $S$ into the frequency domain ($X$) with the help of Fast Fourier Transform (FFT)~\cite{FFT, flydl}, and then the filtering operation is implemented through dot product operation. Finally, the fully denoised sequence representation $\widetilde{S}$ is obtained by inverse transformation. Due to the $O(N \log N)$ computational complexity of the FFT, this denoising process is executed with remarkable efficiency. The specific formula is as follows:

\begin{equation}
X=\mathcal{F}\left(S\right) \in \mathbb{C}^{L_{seq} \times d}, \widetilde{{X}}={W} \odot {X},  \widetilde{S} = \mathcal{F}^{-1}\left(\widetilde{X}\right) \in \mathbb{R}^{L_{seq} \times d},
\end{equation}
where $S$ represents the user behavior sequence, $X$ represents the frequency domain representation of $S$, $\widetilde{S}$ represents the denoised sequence features, $W$ denotes the dot product matrix and $\mathbb{C}$ denotes the complex space.

However, this approach encounters two significant challenges. First, the dot product operation fails to effectively integrate information across various frequency bands of the model. In the time domain, this inadequacy manifests as an inability to comprehensively capture the user's interest information across multiple time scales. Second, the number of parameters in the weight matrix $W$ increases with the length of the input sequences. As a result, the model struggles to adapt flexibly to changes in input length, making it difficult to maintain efficiency and performance with varying sequence sizes.

To address these challenges, we propose the Efficient Filter Layer (EFL) and make enhancements to two key aspects of the dot-product operation involving the $W$ matrix. First, we replace the dot product with matrix multiplication to achieve better fusion of information across different frequency bands. This improves the model's ability to capture user interest information at multiple time scales. However, this change further increases the number of model parameters. To mitigate this issue, we introduce the Chunked Diagonal Mechanism, which allows model parameters to be shared among different tokens, enabling the model to handle extremely long sequences without a significant increase in parameters. The specific process is as follows:

\textit{\textbf{Frequency-Aware Fusion}.} In the EFL, the first improvement involves utilizing matrix multiplication instead of the traditional dot product operation. Specifically, we redefine the weight matrix \(W\) to be in \(\mathbb{C}^{L \times d \times d}\) rather than \(\mathbb{C}^{L \times d}\). This modification enables the effective fusion of user behavioral information across different frequency bands. By capturing user interests at multiple time scales, both long-term and short-term, the EFL facilitates a more comprehensive and efficient mining of user behavior patterns. However, while this method significantly improves frequency domain fusion, it also further increases the number of model parameters. To address this, we further introduce the Chunked Diagonal Mechanism.

\textbf{\textit{Chunked Diagonal Mechanism.}} To address the issue of increasing matrix parameters as user sequences grow, the Efficient Filter Layer (EFL) introduces a Chunked Diagonal Mechanism for complex weight matrices. This mechanism is designed to effectively manage parameter growth while maintaining the ability to fully mine sequences and adapt to different sequence lengths. 
The weight matrix $W \in \mathbb{C}^{L \times d \times d}$ is decomposed into $k$ shared weight matrices $W^{n} \in \mathbb{C}^{d / k \times d / k}$ ($n=1, \ldots, k$), each with reduced dimensions. This decomposition into $k$ smaller diagonal weight matrices, denoted as $W^n$, is somewhat interpretable, similar to $k$-head attention, while enabling computational parallelization. So we get $\tilde{x}_{i}^{n}=W^{n} x_{i}^{n}$, where ${x}_{i}^{n}$ represents the $n$-th block of the $i$-th frequency token ($i \in [1,L//(d/k)]$). Specifically, we employ a double-layer MLP structure as $W^{n}$. The formula is as follows:
\begin{equation}
\tilde{x}_{i}^{n}=\operatorname{MLP}\left(x_{i}^{n}\right)=W_{2}^{n} \sigma\left(W_{1}^{n} x_{i}^{n} +b_{1}^{n}\right)+b_{2}^{n}, 
\end{equation}
where \(W_1^n\) and \(W_2^n\) are the weights, \(b_1^n\) and \(b_2^n\) are the biases, and \(\sigma\) denotes an activation function. Importantly, these weights and biases are shared across all tokens, significantly reducing the overall parameter count. By decomposing the weight matrix into chunked diagonal components and using shared parameters, the model can efficiently handle sequences of different lengths without an increase in parameters. This approach ensures that the model remains scalable and adaptable, maintaining high performance and computational efficiency.

Next, to enhance the model's nonlinear modeling capabilities and stability, we utilize residual connections and Feed-Forward Network (FFN) structures, significantly boosting the model's ability to capture and model user behaviors. Additionally, to further leverage the capabilities of multi-behavior sequential recommendations and understand the relationships between different behaviors, we process the user's various behavior sequences as well as the overall behavior sequence through the Efficient Filter Layer (EFL). This filtering process produces refined representations for each behavior sequence. Subsequently, these filtered representations are passed through a behavior mixer, which is a fully connected structure, to obtain the final representation. The behavior mixer effectively combines information from different behavior sequences, capturing both the individual behaviors and their temporal relationships. This integration enhances the model's ability to make accurate and comprehensive recommendations. The culmination of these layers is the final module, named the Efficient Behavior Miner (EBM). In conclusion, our proposed EBM can effectively capture complex user patterns to filter out data noise and maintain model simplicity by utilizing frequency-aware fusion and the chunking diagonal mechanism.

\begin{table}[h]
\centering
\caption{Complexity, parameter count, and degree of feature fusion for SA, FMLP, and EBM. $L$, $d$, and $k$ refer to the sequence length, hidden size, and block count, respectively.}
\scalebox{0.92}{
\begin{tabular}{l|l|l|l}
\hline
Model         & Complexity                                   & Parameter Count                                                                                       & Feature fusion \\ \hline
Self-Attention & $L^2d+3Ld^2$ &$3d^2$ & Adequate      \\
FMLP          & $Ld+Ld\log L$                             & $Ld$                                                                                                  &   Inadequate      \\
EBM           & $Ld^2/k+Ld\log L$                           &   $(1+4/k)d^2+4d$                                    & Adequate      \\
\hline
\end{tabular}
}
\label{tab:ef}
\end{table}

\textbf{Efficiency Analysis}. To further illustrate the efficient denoisng module EBM, we conduct a complexity comparison analysis with some representative methods, such as Self-Attention and FMLP~\cite{FMLP}, as shown in Table~\ref{tab:ef}. From the table, we can conclude that both EBM and FMLP have lower complexity compared to traditional self-attention methods. Although FMLP is based on Fourier Transform operation to accelerate efficiency, it only utilizes the dot-product operator and fails to fuse frequency domain information, which cannot model more complex patterns in behavior sequences. In contrast, EBM can achieve full fusion (The ability to fully intersect global frequency domain features.) while requiring relatively small parameters, independent of behavior sequence length. This ensures the efficiency of the DPCPL in terms of both running time and parameter spaces.

Through $L$-layer EBMs, we obtain the final user representation $u$. Finally, we define the pre-training loss function. In MBSR, the prediction of the next item under different behaviors is typically the main task. To efficiently migrate to downstream tasks, we design a pre-training task that focuses on predicting both the next item (with indistinguishable behaviors) and the type of next behavior.  For \((u,b_p,v_{p}) \in S^{+}\) and (\(u,b_n,v_{n}) \in S^{-}\), the specific loss functions are presented below:

\begin{equation}\label{eq: loss_pretrain}
\mathcal{L}_{p t}=-\sum_U \sum_S \log \sigma\left(u^T e_{p}-u^T e_{n}\right)-\log \sigma\left(u^{\top}  b_{p}-u^{\top}  b_{n}\right),
\end{equation}
where \(S^+\) represents the positive sample pairs, and \(S^- \) represents the negative ones. Specifically, $e_{p}$ represents the item embeddings for positive samples, which are items the user has interacted with, while \(e_{n}\) denotes the embeddings of negatively sampled items that are randomly selected.

\subsection{Customized Prompt Tuning}
% After pre-training, we intend to utilize the prompt tuning technique to migrate knowledge to downstream tasks for efficiency. However, unlike NLP , the tokens of the pre-trained recommendation model lack semantic information, rendering it difficult to manually design hard prompts. So we leverage the pre-trained model in the form of soft prompts. To fully leverage the pre-trained model, we intend to encourage three required properties of prompts: \textbf{Personalization}, \textbf{Progressiveness}, and \textbf{Diversity}.

After pre-training, we aim to utilize prompt tuning techniques to transfer knowledge to downstream tasks efficiently. It is worth mentioning here that compared to very early user behavior, the data we use for fine-tuning is the user's behavior data in the most recent period. However, unlike in NLP contexts~\cite{unlikenlp1, unlikenlp2, unlikenlp3, q2}, the tokens of a pre-trained recommendation model lack semantic information, making it challenging to manually design effective hard prompts. Specifically, in recommendation models, the tokens are item IDs, which do not carry specific semantics. This absence of inherent meaning makes it difficult to create hard prompts based on grammatical rules as done in NLP. Moreover, personalization is crucial in recommendation systems, requiring prompts to be tailored to individual users. Therefore, we propose that prompt learning in recommendation scenarios must be personalized and introduce a new concept called Customized Prompt Learning. This approach involves dynamically setting prompts based on various user-specific auxiliary information.

To leverage the pre-trained model effectively, we use soft prompts. Our goal is to ensure these prompts possess three key properties: Personalization, Progressiveness, and Diversity. By doing so, we aim to maximize the pre-trained model's potential and enhance the efficiency and accuracy of downstream tasks in recommendation systems.

\subsubsection{Prompt Personalization}
To explore the personalized behavioral patterns in the pre-trained model, we propose the concept of Customized Prompt Learning to personalize prompt learning for each user. Specifically, we dynamically customize the prompts based on the user's personalized information, whereas, in MBSR, we can take into account the user's attributes, statistics, and behaviors. 

First, as for attributes and statistics, we embed them directly. It is worth mentioning that in MBSR, we found and firstly pointed out that statistics is a crucial feature. For instance, the number of various behaviors of a user can reflect whether the user is in a cold-start group, and the conversion ratio between behaviors (e.g., the proportion of purchases made by adding to a cart) can reflect the user's behavioral habits. Second, in order to model the variability of different behaviors, we model the different sequences of users' behaviors separately to generate prompts through a simple yet efficient gated recurrent network. Formally, for user \(u\), we get prompts about the user's attributes \({q}_{u,attr}\) through \(MultiMLP([{a}_{1}^{u}\left\|{a}_{2}^{u}\right\| \cdots \| ])\), and with the same token we can get prompts about the user's statistics \({q}_{u,statis}\), where the $||$ symbol represents vector concatenation. In addition, we represent the user's behavioral prompt information \({q}_{u,b}\) by the hidden state of the last layer of the Gated Recurrent Unit (GRU) at the final time step. 

\subsubsection{Prompt Progressiveness}
Beyond personalization, we aim to further exploit the potential of pre-trained models by designing elaborate prompt strategies. To achieve this, we find that the pre-trained model uses a bottom-up learning process, where user interests are learned through a step-by-step hierarchical condensation process of user behaviors. Considering this, we develop an innovative progressive prompt strategy. Specifically, our progressive prompt approach incorporates prompts not only at the input side of the model but also at each layer of the model. This approach enables prompts to steer and enhance the model's comprehension of user preferences as it advances through each layer, enabling more efficient fine-tuning of pre-trained models and improving performance on downstream tasks such as cold-start. 

Further, we considered that the various layers of the model require distinct prompt factors, which are obtained from the user's personalized information. Extracting the appropriate prompt factors for each layer from the complex user information poses a significant challenge. To this end, we propose a Prompt Factorized Gate Network (PFG) structure for refining the prompt factors from the prompt information and then generating the prompts for each layer of the pre-trained model.

Specifically, following the personalized prompt generator, we acquire various aspects of prompt information referred to as \(Q_u\) for user \(u\). Initially, we consolidate this prompt information into a prompt factor matrix of size $L \times N$, where \(L\) is the number of network layers and \(N\) is the number of prompt factors in each layer through PFG. The specific formula used is as follows:
\begin{equation}
Q_u=[q_{u,attr}|| q_{u,statis}||...||q_{u,b}],
\end{equation}
\begin{equation}\label{eq: layer_wise_prompt}
\mathcal {A}^u_{l,n}=\mathrm{softmax}\left(W_{l,n} Q_u\right), \quad  E_{l,n}^u = \mathcal {A}^u_{l,n} Q_{u}, 
\end{equation}
where the $||$ symbol represents vector concatenation.

In addition, we believe that some prompts may be shared in different processes of model learning. To this end, we use the same method to generate a set of shared prompt factors \(E_{s,n}^u\). Therefore, for each layer of the model, we employ the shared and layer-specific factors as input to generate the corresponding layer's prompt through the Prompt Factorized Gate Network structure. The specific formula is as follows: 
\begin{equation}
\Phi_l^u = [E_{l,1}^u||E_{l,2}^u,...,E_{l,N}^u||E_{s,1}^u||E_{s,2}^u,...,E_{s,N}^u],
\end{equation}
\begin{equation}
\beta_{j}^u=\mathrm{softmax}\left(W_l \Phi_u\right), \quad p_{l}^u = \left(\sum_{j=1}^{2N} \beta_{j}^u E_j^u\right),
\end{equation}

\begin{equation}
p_{l,c}^{u} = W_{l,c}p_{l}^{u}.
\end{equation}

Finally, we obtain the prompt \(p_{l}^u\) for user \(u\) employed within layer \(l\) of the pre-trained model and place the prompt in the first \(C\) tokens which are analyzed in the hyperparameter experiment.

\subsubsection{Prompt Diversity}
However, the prompts of different layers in Eq. (\ref{eq: layer_wise_prompt}) tend to be easily assimilated, because the information sources generating the prompts are consistent. Although we designed the PFG to fully decouple and fuse the prompt information, it is still challenging to guarantee the process of prompt generation is controllable. To further ensure the diversity of prompts, we introduce a regularization loss based on representation compactness~\cite{compactness}. 

Compactness is a desired trait of intra-factor representations and its opposite is what we expect for inter-factor representations. ReduNet \cite{ReduNet} proposed to measure compactness of representation with rate-distortion \(R(z, \epsilon)\), which determines the minimal number of bits to encode a random variable \(z\) subject to a decoding error upper bounded by \(\epsilon\). Inspired by this, we design a compactness regularization loss function to control the diversity of the prompt vector space. Ideally, diversity should be maintained between prompt factors, and between prompts across layers for each user. The specific formula is as follows:
\begin{equation}
R(\mathcal{E}, \epsilon)=\frac{1}{2} \log \mathrm{det}\left(I+\frac{d}{N \epsilon^{2}} \mathcal{E} \mathcal{E}^{T}\right),
\end{equation}

\begin{equation}
R(\mathcal{P}, \epsilon)=\frac{1}{2} \log \mathrm{det}\left(I+\frac{d}{L \epsilon^{2}} \mathcal{P} \mathcal{P}^{T}\right),
\end{equation}
where $\mathcal{E} \in \mathbb{R}^{|N\times (L+1)| \times d}$ is the prompt factors matrix, $\mathcal{P} \in \mathbb{R}^{|L| \times d}$ is the prompt matrix, and the rest are hyperparameters. $logdet(
\cdot)$ means the logarithm of the determinant of a matrix and $I$ is the identity matrix. Finally, we can obtain our compactness regularization loss as follows:
\begin{equation}
\mathcal{L}_{ {compactness }}=\sum_{u_i\in U}[\lambda_e R(\mathcal{E}_i, \epsilon_e) + \lambda_p R(\mathcal{P}_i, \epsilon_p)].
\end{equation}
% The \(\mathcal{L}_{pred}\) function exhibits similarities to \(\mathcal{L}_{pt}\), thus avoiding redundancy in this paper. The only difference the model predicts is the interaction item under the target behavior. Consequently, the total loss of the prompt-tuning stage can be derived as follows:
Through the compactness regularization loss, we ensure diversity in prompt factors and the final prompt vector across different layers. This guarantees that prompts from the same personalized information at different layers meet the criteria of progression. This is because it reinforces the understanding that prompts at different layers are distinct, aligning with the assumption of continuous refinement of interests across the model's layers.
Consequently, the total loss of the prompt-tuning stage can be derived as follows:
\begin{equation}
\mathcal{L}_{total} = \mathcal{L}_{pred} + \lambda \mathcal{L}_{compactness},
\end{equation}
where \(\lambda\) is a hyperparameter used to control the strength of compactness regularization, and \(\mathcal{L}_{pred}\) is similar to \(\mathcal{L}_{pt}\) in Eq. (\ref{eq: loss_pretrain}) but only predicts item under a given target behavior.

By employing the three properties of prompts—\textbf{Personalization}, \textbf{Progressiveness}, and \textbf{Diversity}, we can obtain a fine-tuned model capable of handling downstream tasks effectively. Due to this paradigm, we can efficiently model long sequences of users' historical behaviors and benefit from a wide range of users' behaviors without significantly increasing the complexity of the model. Subsequent experiments also proved that the effect of fine-tuning is equal to or even exceeds the effect of full-scale fine-tuning of the model.

\subsection{Essential Analysis of prompts}

Finally, we delve into the essence of prompt and the reasons why customized prompt learning is effective in pre-trained recommendation models without semantics. \textbf{Firstly, the essence of prompt is a kind of instruction to the task.} In NLP, for the same task, the same prompt can be used for different inputs because the token in NLP is highly semantic; however, in recommendation, the token is just the ID of the item, which is not semantic. For this reason, we propose the concept of customized prompt learning, i.e., generating personalized prompts for different users, which is equivalent to giving different commands to the model for different users, enhancing the personalization of the model's output and improving the recommendation effect.
\textbf{Secondly, the essence of prompt is a kind of pre-training task reuse.} Another reason for the effectiveness of our approach is that we design the next prediction pre-training task, as well as generating prompts with pre-trained historical user behaviors, trying to make the fine-tuning task reuse the pre-training task, and avoiding the introduction of a large number of new parameters that would lead to catastrophic forgetting of the model.
\textbf{Thirdly, the essence of the prompt is searching for valid parameters.} The goal of prompt-tuning, from fixed templates to learnable templates, as well as the wide variety of tuning strategies proposed, is to achieve a better adaptation of the model while tuning the model with as few parameters as possible. So prompt learning is about finding the part of the parameters that are more effective for model adaptation. Just as we analyzed that the model is going layer by layer to condense the learning interests from the behaviors, we propose to add some prompts to each layer of the pre-trained model, which will have a more direct impact on the model in learning user interests, and one of the reasons why DPCPL is effective is that we have found the part of the model parameters that are more effective in the model adaptation.

\section{EXPERIMENT}
\subsection{Experimental Setting}

\subsubsection{Datasets}

\begin{table}[]
%\begin{minipage}{\textwidth}
\caption{Statistical information of experimented datasets.}
\resizebox{\linewidth}{!}{

\begin{tabular}{ccccc}
\hline
Dataset        & CIKM                                                                              & IJCAI                                                           & Taobao     &                      \\ \hline
\#users        & 254,356                                                                           & 324,859                                                         & 279,052    &                      \\
\#items        & 521,900                                                                           & 331,064                                                         & 731,517    &                      \\
\#interactions & 31,824,670                                                                        & 46,694,666                                                      & 36,758,555 &            
       \\
\#Density & 	2.4e-4                                                                       & 	4.3e-4                                                      & 1.9e-4 &            
       \\
\#Avg-length & 	125.12                                                                        & 143.74                                                      & 	136.62
 &            
       \\
User profile   & \begin{tabular}[c]{@{}c@{}}gender, age, \\ consumption\end{tabular} & \begin{tabular}[c]{@{}c@{}}gender,\\ age\_range\end{tabular} & None       & \multicolumn{1}{l}{} \\ \hline

\end{tabular}
}

\label{tab:data}

\end{table}
%\end{minipage}

\begin{table*}
\centering
\small
\caption{Overall performance comparison of all methods in terms of HR@K and NDCG@K. (p-value \(<\) 0.05)}

\begin{tabular}{c|cccc|cccc|cccc}
\hline
Datasets             & \multicolumn{4}{c|}{CIKM}                                                                                                                     & \multicolumn{4}{c|}{Taobao}                                                                                                                                                                    & \multicolumn{4}{c}{IJCAI}                                                                                                                     \\ \hline
Metric               & {H@10} & {N@10} & {H@20} & {N@20} & {H@10} & {N@10} & {H@20} & {N@20} & {H@10} &{N@10} & {H@20} & {N@20} \\ \hline
GRU4Rec              & 0.270     & 0.155     & 0.311     & 0.179     & 0.245     & 0.139     & 0.268     & 0.155     & 0.296     & 0.164     & 0.365     & 0.208     \\
SASRec               & 0.325     & 0.198     & 0.423     & 0.243     & 0.318     & 0.188     & 0.408     & 0.226     & 0.373     & 0.200     & 0.484     & 0.249     \\
Bert4Rec             & 0.356     & 0.228     & 0.450     & 0.248     & 0.304     & 0.178     & 0.391     & 0.212     & 0.398     & 0.219     & 0.520     & 0.283     \\ 
FEARec             & 0.363     & 0.229     & 0.463     & 0.259     & 0.324     & 0.190     & 0.416     & 0.227     & 0.409     & 0.223     & 0.533     & 0.285     \\ 	 	 	 	 	 	 	 	 	
\hline
MBGCN                & 0.364     & 0.227     & 0.467     & 0.263     & 0.333     & 0.196     & 0.427     & 0.235     & 0.412     & 0.224     & 0.536     & 0.284     \\
MBHT & 0.360     & 0.223     & 0.466     & 0.265     & 0.342     & 0.201     & 0.438     & 0.242     & 0.409     & 0.202     & 0.524     & 0.279     \\

KGHT &0.380 	&0.228 	&0.489 	&0.280 	&0.357 	&0.207 	&0.454 	&0.251 	&0.411 	&0.222 	&0.534 	&0.280 \\
CML	&0.369 	&0.229 	&0.476 	&0.269 	&0.345 	&0.203 	&0.442 	&0.244 	&0.420 	&0.227 	&0.546 	&0.287    \\
CKML	&0.399 	&0.233 	&0.512 	&0.295 	&0.373 	&0.213 	&0.470 	&0.260 	&0.410 	&0.222 	&0.534 	&0.282    \\
KMCLR                 & 0.413     & 0.227     & 0.526     & 0.308     & 0.386     & 0.214     & 0.479     & 0.265     & 0.385     & 0.209     & 0.502     & 0.266     \\ \hline
MB-S\(^3\)Rec        & 0.427     & 0.245     & 0.549     & 0.318     & 0.406     & 0.232     & 0.512     & 0.284     & 0.436     & 0.235     & 0.567     & 0.297     \\
MB-CL4SRec            & 0.409     & 0.248     & 0.522     & 0.296     & 0.427     & 0.244     & 0.538 & 0.298     & 0.437     & 0.238     & 0.570     & 0.304     \\
MBPPR                & 0.448     & 0.257     & 0.576     & 0.334     & 0.370     & 0.214     & 0.471     & 0.258 & 0.458     & 0.247     & \underline{0.616} & 0.312     \\
CPTPP                & 0.428     & 0.252     & 0.549     & 0.315     & \underline{0.449} & \underline{0.256} & \underline{0.565} & \underline{0.313} & 0.447     & 0.243     & 0.593 & 0.304     \\
DPT                  & \underline{0.471} & \underline{0.270} & \underline{0.606} & \underline{0.351} & 0.398     & 0.229     & 0.504     & 0.278 & \underline{0.482} & \underline{0.260} & 0.607     & \underline{0.328} \\ \hline
\textbf{DPCPL} & \textbf{0.519} & \textbf{0.299} & \textbf{0.664} & \textbf{0.383} & \textbf{0.484} & \textbf{0.275} & \textbf{0.609} & \textbf{0.336} & \textbf{0.523} & \textbf{0.283} & \textbf{0.669} & \textbf{0.359} \\
Improvement          & 10.20\%    & 10.44\%    & 9.63\%     & 9.18\%     & 7.91\%     & 7.38\%     & 7.79\%     & 7.41\%     & 8.56\%     & 9.06\%     & 8.57\%     & 9.72\%     \\ \hline

\end{tabular}

\label{tab:op}

\end{table*}

To comprehensively investigate the performance of DPCPL, we conduct experiments on three large-scale real-world recommendation datasets, which are widely utilized in multi-behavior sequential recommendation research and are considered standard benchmarks \cite{MBHT, MBGMN}. These user behavior datasets contain various interactions: click, add-to-cart, add-to-favorite, and purchase. 

\textbf{CIKM}\footnote{https://tianchi.aliyun.com/dataset/35680}. The CIKM dataset is an E-commerce recommendation dataset that was released by Alibaba. It is specifically designed to aid in the development and evaluation of recommendation algorithms within the E-commerce domain. 

\textbf{IJCAI}\footnote{https://tianchi.aliyun.com/dataset/42}. This dataset is released by IJCAI Contest 2015 for the task of repeat buyers prediction. This dataset provides extensive data on user purchasing behavior, which can be leveraged to develop and test algorithms aimed at identifying customers who are likely to make repeat purchases. 

\textbf{Taobao}\footnote{https://tianchi.aliyun.com/dataset/649}. This dataset is collected from Taobao, which is one of the largest e-commerce platforms in China. This dataset offers a wealth of information on user interactions and transactions on the platform, making it an excellent resource for developing and testing various E-commerce applications, such as recommendation systems, user behavior analysis, and sales forecasting.

In data processing, the experiment aims to maintain the original dataset's distribution to reflect real-world characteristics accurately. We filter out user and item data with fewer than 20 interactions. The first 60\% of the data, based on timestamps, is designated as pre-training data, while the remaining 40\% is used for fine-tuning and validation tests. For non-pre-trained models, no pre-training data is utilized in the experiments. This is because recommendation algorithms that often input sequence lengths that are limited, and generally only input the user’s most recent behavior. The pre-training fine-tuning paradigm can mitigate this issue to some extent. It allows for pre-training with ultra-long historical behaviors and fine-tuning using the most recent behavioral data. Detailed statistical information about these datasets is summarized in Table~\ref{tab:data}.

\subsubsection{Comparison Baselines} We compare DPCPL with various state-of-the-art baseline methods, including general sequential recommendation, multi-behavior sequential recommendation, and pre-training fine-tuning approaches.

\textbf{General Sequential Recommendation Methods.}
In this context, the sequential recommendation algorithm inputs only the items with which the user interacts, without considering the type of behavior.

\textit{\textbf{GRU4Rec}} \cite{GRU4Rec}. It utilizes the gated recurrent unit as sequence encoder to learn dynamic preference.

\textit{\textbf{SASRec}} \cite{SASRec}. In SASRec, self-attention is employed to encode the sequential patterns of user interaction, without the recurrent operations over input sequences.

\textit{\textbf{BERT4Rec}} \cite{BERT4Rec}. It uses a bidirectional encoder for modeling sequential information with transformer, which is optimized with the Cloze objective and has produced state-of-the-art performance among many baselines.

\textit{\textbf{FEARec}} \cite{FEARec}. It improves the original time domain self-attention in the frequency domain with a ramp structure, allowing for the explicit learning of both low-frequency and high-frequency information. 

\textbf{Multi-Behavior Recommendation Methods.} In the multi-behavior recommendation algorithm, we take into account the relationships between different behavior types.

\textit{\textbf{MB-GCN}} \cite{MBGCN}. It models the multi-behavior
of users and uses graph convolutional network to perform
behavior-aware embedding propagation.

\textit{\textbf{MBHT}} \cite{MBHT}, which transform multi-behavior interactions into a unified graph or hyper-graph.

\textit{\textbf{KHGT}} \cite{KHGT}. It introduces a transformer-based approach to multibehavior modeling with emphasis on temporal information and auxiliary knowledge information, modeling behavior embeddings through graph attention networks.

\textit{\textbf{CML}}~\cite{CML}. It introduces CL into multi-behavior recommendation, proposing meta-contrastive coding to enable the model to learn personalized behavioral features.

\textit{\textbf{CKML}}~\cite{CKML}, which uses Coarse-grained Interest Extracting (CIE) and Fine-grained Behavioral Correlation (FBC) modules, along with a self-attention mechanism, to capture and correlate shared and behavior-specific interests.

\textit{\textbf{KMCLR}}~\cite{KMCLR} which enhances recommender systems through two comparative learning tasks and three functional modules: multiple user behavior learning, knowledge graph enhancement, and coarse- and fine-grained modeling of user behaviors to improve performance.

\textbf{Pre-trained and Prompt-tuning Methods}. In the pre-training method paradigm, we can consider incorporating the user's long-term historical behavior sequence for pre-training. This would be followed by fine-tuning using the user's recent behavior sequence with a typical sequence length, just like the non-pre-training method. It is worth noting that we follow the relevant literature \cite{MBHT} and transform the original non-multi-behavior method into a multi-behavior sequence recommendation method MB-X by adding behavior embedding.

\textit{\textbf{S\(^3\)Rec}} \cite{S3Rec} uses SSL with a pre-training strategy to derive the intrinsic data correlation. In this section, we only compare the mask item prediction (MIP) and sequence-segment correlation segment prediction (SP) in S\(^3\)Rec for fairness. 

\textit{\textbf{CL4SRec}} \cite{CL4SRec} combines contrastive SSL with a Transformer-based SR model. The model incorporates crop, mask, and reorder augmentation operators. 

\textit{\textbf{PPR}} \cite{PPR} represents a groundbreaking contribution that employs prompt learning to address the challenge of the cold start problem by generating prompts solely based on user attribute information. 

\textit{\textbf{CPTPP}} \cite{CPTPP}. 
The CPTPP framework leverages user information and graph-based contrastive learning to effectively align pre-trained user vectors with specific downstream tasks, enhancing performance by capturing intricate user relationships and preferences.

\textit{\textbf{DPT}} \cite{DPT}  adopts a three-stage denoising and a fine-tuning approach to reduce noise impacts.

\subsubsection{Implementation Details}
To ensure fair comparisons, we maintain consistent settings across all methods. The embedding size is set to 256, and all methods incorporate the embedding of user attribute information. During the experimentation process, we perform a grid search to identify the optimal hyperparameters. The maximum number of epochs is set to 1000, and training is halted if the NDCG@K summation on the validation dataset does not improve for 20 consecutive epochs.
For our proposed method, the ratio of the compactness loss \(\lambda\) is tuned among \(\{1 \times 10^{-3}, 1 \times 10^{-2}, 1 \times 10^{-1}\}\), and the value of \(\epsilon^2\) is tuned among \(\{0.5, 1, 5\}\). As for the baselines, we tune their hyperparameters according to the guidelines provided in their original papers and initialize these parameters using Xavier Initialization~\cite{xcsh}. By adhering to these consistent settings and carefully tuning the hyperparameters, we aim to ensure that our comparisons are both rigorous and fair, providing reliable insights into the performance of the proposed method relative to established baselines.

\subsubsection{Evaluation Metrics}
In this study, we assess the performance of comparison methods for the top-K recommendation using two evaluation metrics: Hit Ratio (HR@K) and Normalized Discounted Cumulative Gain (NDCG@K). HR@K measures the average proportion of relevant items in the top-K recommended lists, while NDCG@K evaluates the ranking quality of the top-K lists in a position-wise manner. For fair and efficient evaluation, each positive instance is paired with randomly selected 100 non-interactive items~\cite{KHGT, MBHT, MBGMN}. In our experiments, we adopt the leave-one-out strategy for performance evaluation. We begin by treating the purchasing as a target behavior and for each user, we regard the temporally ordered last purchase as the test sample and the previous ones as validation samples.

\subsection{Overall Performances}

In Table~\ref{tab:op}, we present a detailed performance comparison across various datasets. The key observations from this analysis are summarized as follows:

(1). We observe that more advanced multi-behavioral approaches tend to achieve more promising results. This demonstrates the effectiveness of incorporating multiple types of behavioral data into sequential modeling. By leveraging richer user multi-behavior information, these models can better understand and predict user preferences.

(2). Compared to end-to-end approaches, pre-trained models consistently outperform, highlighting the potential of the pre-training paradigm to enhance downstream task performance. Pre-training empowers models to learn from extensive datasets, capturing intricate patterns that can be fine-tuned for specific tasks, resulting in superior performance. Moreover, as the number of parameters in a pre-trained model increases, its memory capacity expands, facilitating the absorption of larger historical datasets and mitigating concerns related to excessively lengthy sequences.

(3). We find that models incorporating denoising techniques during pre-training outperform other methods. This suggests that historical user behavioral noise significantly impacts model performance and needs to be addressed. By removing noise from historical data, these models can learn more accurate representations of user behaviors.

(4). Our proposed DPCPL consistently outperforms other methods across all datasets. This highlights the effectiveness of our customized prompt learning and progressive prompt strategies. By dynamically setting prompts based on user-specific information and progressively refining these prompts, DPCPL can better adapt to individual user behaviors and improve recommendation accuracy.

\begin{table}[]
\small
\caption{Efficiency experiments compared to fine-tuning.}

\resizebox{\linewidth}{!}{
\begin{tabular}{c|c|cc|cc}
\hline
Datasets                 & Metric                            & Fine-tuning & DPCPL & Para                     & Time                      \\ \hline
\multirow{2}{*}{CIKM}   & H@10 & 0.526       & 0.519   & \multirow{2}{*}{1.25\%} & \multirow{2}{*}{8.15\%}  \\
                         & N@10 & 0.302       & 0.299   &                          &                           \\ \hline
\multirow{2}{*}{Taobao} & H@10 & 0.494       & 0.484   & \multirow{2}{*}{1.42\%} & \multirow{2}{*}{12.96\%} \\
                         & N@10 & 0.281       & 0.275   &                          &                           \\ \hline
\multirow{2}{*}{IJCAI}  & H@10 & 0.480       & 0.523   & \multirow{2}{*}{0.86\%} & \multirow{2}{*}{7.43\%}  \\
                         & N@10 & 0.260       & 0.283   &                          &                           \\  \hline
\end{tabular}
}

\label{tab:ff}

\end{table}

\begin{table}
\centering
\caption{Ablation study with key modules.}

\resizebox{\linewidth}{!}{
\begin{tabular}{c|cc|cc|cc}
\hline
Datasets         & \multicolumn{2}{c|}{CIKM}                                             & \multicolumn{2}{c|}{Taobao}   
& \multicolumn{2}{c}{IJCAI} \\ \hline
Metric           & {H@10} & {N@10} & {H@10} & {N@10} & {H@10} & {N@10}\\ \hline
w/o DS      & 0.457                             & 0.262                             & 0.430                             & 0.251         & 0.444     & 0.241                \\
w/o PS           & 0.424                             & 0.243                             & 0.404                             & 0.235           & 0.414     & 0.221              \\
w/o PG           & 0.478                             & 0.263                             & 0.455                             & 0.251        & 0.470     & 0.266             \\
w/o CT           & 0.495                             & 0.284                             & 0.471                             & 0.269           & 0.496     & 0.255           \\ \hline

\textbf{DPCPL} & \textbf{0.519}                    & \textbf{0.299}                    & \textbf{0.484}                    & \textbf{0.275}              & \textbf{0.523}   & \textbf{0.283}  \\ \hline
\end{tabular}
}

\label{tab:as}

\end{table}

\subsection{Analysis of Efficiency}

To explore the efficiency of customized prompts learning, we compare the time overhead and number of parameters with full parameters fine-tuning. Surprisingly, it even approaches or surpasses the overall fine-tuning effect to some extent. In addition, compared to full fine-tuning, DPCPL requires only a limited number of parameters to be tuned and stored, thus demonstrating parameter efficiency. As indicated in Table~\ref{tab:ff}, the parameter update count in DPCPL amounts to only 1.2\%, 1.4\%, and 0.8\% of that required for full fine-tuning for each dataset. Accordingly, the time required for DPCPL is merely 8.15\%, 12.96\%, and 7.43\% of that needed for full fine-tuning, respectively. It is important to note that since most baselines are based on transformer architectures and typically require full fine-tuning, their performance results are similar to the full fine-tuning method. Compared to the prompt-based fine-tuning method, the robust baseline model DPT, our approach achieves substantial reductions in parameter count and fine-tuning time when applied to the CIKM dataset. Specifically, our method decreases the parameter count to 34.6\% and reduces the fine-tuning time to 23.5\% of DPT's requirements. Moreover, our approach demonstrates improvements of approximately 10\% across multiple experimental metrics.

In addition to the high efficiency of parameters and time adjusted during the fine-tuning phase, our pre-trained model is also very efficient in the previous complexity analysis in Table~\ref{tab:ef}. It is noteworthy that the EBM module maintains a time complexity of \textit{$Ld^2/k+Ld\log L$} and a parameter count of \textit{$(1+4/k)d^2+4d$}, which is substantially lower than that of self-attention mechanisms. This design ensures efficient handling of behavioral sequences, with the added benefit of a parameter size that is agnostic to the length of the sequence.

In conclusion, our pre-trained model architecture follows a linear structure, providing a solid foundation in terms of parameter count and time complexity. Additionally, through customized prompt fine-tuning, we significantly reduce the number of parameters involved in fine-tuning, resulting in faster processing. This highlights that our DPCPL not only delivers impressive results but also demonstrates high efficiency in both pre-training and fine-tuning processes.

\subsection{Ablation Studies}
To investigate the effectiveness of each component, we introduce the following variants of DPCPL:

(1).  \textit{\textbf{Variant without Denoising Block (w/o DS)} } Instead of the EBM block, we exclusively employ the attention block in the pre-trained model.

(2).  \textit{\textbf{Variant without Personalized Prompts (w/o PS)} } In this variant, instead of employing personalized prompts, we utilize continuous (soft) prompts, which are randomly initialized and trainable, drawing inspiration from methodologies in the field of NLP.

(3).  \textit{\textbf{Variant without Progressive Prompts (w/o PG)}} Instead of progressive prompts, we only incorporate prompts into the first layer of our model.

(4).  \textit{\textbf{Variant without Controllability (w/o CT)}} In this variant, we do not consider the controllability of prompts. Thus, we omit the regularization loss.

The results are presented in Table~\ref{tab:as}. 
Firstly, the (w/o DS) variant demonstrates the effectiveness of our proposed Efficient Behavior Miner (EBM), which can efficiently filter out the noise in the user's multi-behavioral sequences in the frequency domain, so as to facilitate the model to better learn the user's behavioral patterns.
Secondly, the (w/o PS) variant demonstrates the validity of our proposed paradigm of customized prompt learning. It highlights the challenge of directly applying the NLP prompt learning paradigm to recommendation problems, necessitating personalized generation. Thirdly, the (w/o PG) variant underscores the significance of progressive prompts. It recognizes that pre-trained models undergo interest condensation in a hierarchical manner, thereby the models require different prompt factors at different layers. This demonstrates the superiority of our proposed Prompt Factor Gate (PFG), which efficiently extracts and combines prompt factors from prompt information. Lastly, the (w/o CT) variant demonstrates the superiority of compactness regularization, which controls the diversity of prompt factors and layer-specific prompts, thereby expanding the representation space extensively.

\subsection{Analysis of the Adaptation to downstream tasks}

\begin{table}[h]
\caption{Adaptation experiments with target behavior cart}

\resizebox{\linewidth}{!}{
\begin{tabular}{c|cc|cc|cc}

\hline
Datasets       & \multicolumn{2}{c|}{CIKM}                                             & \multicolumn{2}{c|}{Taobao}                                           & \multicolumn{2}{c}{IJCAI}                                             \\ \hline
Metric         & {{H@10}} & {N@10} & H@10 & N@10 & {H@10} & N@10 \\ \hline
MBPPR      & 0.404                             & 0.233                             & 0.367                             & 0.208                             & 0.444                             & 0.246                             \\
DPT          & 0.487                             & 0.297                             & 0.477                             & 0.282                             & 0.559                             & 0.299                             \\
CPTPP            & 0.534                             & 0.342                             & 0.455                             & 0.266                             & 0.598                             & 0.328                             \\ \hline
\textbf{DPCPL} & \textbf{0.582}                    & \textbf{0.359}                    & \textbf{0.499}                    & \textbf{0.293}                    & \textbf{0.638}                    & \textbf{0.345}                    \\ \hline

\end{tabular}

}

\label{3}
\end{table}

\begin{table}[h]

\caption{Adaptation experiments with target behavior click.}

\resizebox{\linewidth}{!}{

\begin{tabular}{c|cc|cc|cc}
\hline
Datasets       & \multicolumn{2}{c|}{CIKM}                                 & \multicolumn{2}{c|}{Taobao}                               & \multicolumn{2}{c}{IJCAI}                                 \\ \hline
Metric         &  H@10 &N@10 &  H@10 &  N@10 &  H@10 & N@10 \\ \hline
MBPPR      & 0.485                       & 0.280                       & 0.440                       & 0.250                       & 0.533                       & 0.295                       \\
DPT          & 0.585                       & 0.356                       & 0.573                       & 0.338                       & 0.671                       & 0.359                       \\
CPTPP            & 0.641                       & 0.411                       & 0.546                       & 0.320                       & 0.717                       & 0.394                       \\ \hline
\textbf{DPCPL} & \textbf{0.677}              & \textbf{0.429}              & \textbf{0.599}              & \textbf{0.352}              & \textbf{0.762}              & \textbf{0.426}              \\ \hline
\end{tabular}
}

\label{2}
\end{table}

\begin{table}[h]
\caption{Adaptation experiments with the cold-start user.}

\resizebox{\linewidth}{!}{

\begin{tabular}{c|cc|cc|cc}
\hline
Datasets       & \multicolumn{2}{c|}{CIKM}                                 & \multicolumn{2}{c|}{Taobao}                               & \multicolumn{2}{c}{IJCAI}                                 \\ \hline
Metric         & {H@10} & {N@10} & {H@10} & N@10 & H@10 &N@10 \\ \hline
MBPPR      & 0.137                       & 0.079                       & 0.125                       & 0.071                       & 0.151                       & 0.084                       \\
DPT          & 0.214                       & 0.131                       & 0.210                       & 0.124                       & 0.246                       & 0.132                       \\
CPTPP            & 0.182                       & 0.116                       & 0.155                       & 0.091                       & 0.203                       & 0.112                       \\ \hline
\textbf{DPCPL} & \textbf{0.310}              & \textbf{0.194}              & \textbf{0.270}              & \textbf{0.158}              & \textbf{0.344}              & \textbf{0.189}              \\ \hline
\end{tabular}

}

\label{1}
\end{table}

In this subsection, we aim to verify the efficient adaptation of our pre-trained model to downstream tasks by conducting experiments on additional tasks. 

The target behavior pertains to the items that have been interacted with within the behavior we endeavor to forecast. As shown in Table ~\ref{3} and Table ~\ref{2}, we add shopping cart and clicks to the target behaviors and compare them with other baselines respectively, and find that DPCPL not only achieves efficient adaptation with the pre-trained model, but also achieves good results on various tasks. 

In addition, we define users with target behaviors less than or equal to 2 as cold-start users and use some of the user data for validation. From table \ref{1}, we observe that for cold-start users, we improve 30\% over the current state-of-the-art model, which proves the effectiveness of DPCPL in small sample scenarios. Compared to traditional multi-behavioral approaches, DPCPL proves to be better at transferring general knowledge to user representations in the presence of insufficient training data, which underscores the generalization ability of our denoising pre-trained model, as well as the knowledge transfer capabilities of customized prompt learning.

\begin{figure*}[h]\centering
	\includegraphics[width=0.8\textwidth]{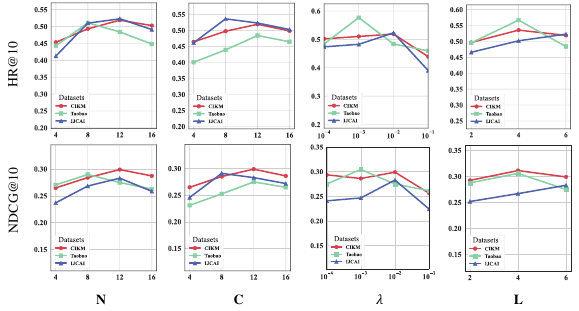}

	\caption{Hyperparameter sensitivity analysis for the selected hyperparameters: number of prompt factors of each group (\(N\)), number of tokens in prompt (\(C\)), compactness regularity parameter (\(\lambda\)), and number of pre-trained model layers (\(L\)).}
    \label{FIG_zhexian}
  
\end{figure*}

\subsection{Hyperparameter Analysis}

To examine the impact of various hyperparameters on DPCPL, we conducted experiments with different configurations of key parameters. In this section, we present the results of these experiments, as depicted in Figure~\ref{FIG_zhexian}. Our conclusions are summarized below:

(1).  \textit{\textbf{Number of prompt factors of each group ($N$).}} We observed that $N$ values between 8 to 12 yielded optimal results. Further increasing the $N$ value seems likely to have a bad effect on the representation possibly due to overfitting, potentially due to the introduction of noise resulting from excessive $N$ values. 

(2).  \textit{\textbf{Number of tokens in prompt ($C$).}} We observed that auxiliary information can be effectively introduced when the token count ranges between 8 to 12. However, increasing the token count beyond this range resulted in a decrease in performance. This decline is potentially due to a reduction in the influence of user behavior input when the token count becomes excessively large, leading to a dilution of the critical information needed for accurate recommendations.

(3).  \textit{\textbf{Compactness regularity parameter ($\bm{\lambda}$).}} $\lambda$ controls the weight of the regularization loss. Observations indicate that $\lambda$ values between 0.01 and 0.001 yield better results. Too much loss of regularization can negatively affect the normal representation of vectors, while insufficient regularization may not add sufficient constraints on the representation of different layers of the prompt. 

(4).  \textit{\textbf{Number of pre-trained model layers ($L$).}} We further study whether the design of the depth feature extraction network is conducive to the recommendation task. We observe that in most cases, the superimposed four-layer network can show better performance, and with the number of network layers increasing, there follows the risk of overfitting. In particular, in the IJCAI dataset, the deeper network achieves better results. The reason for this phenomenon may be that the IJCAI dataset has a larger volume of data and a longer average length than other datasets CIKM and Taobao, which indicates the necessity of the deep network in processing larger and more complex datasets.

\subsection{Visualization Analysis.}
\begin{figure}
    \centering
    \includegraphics[width=0.47\textwidth]{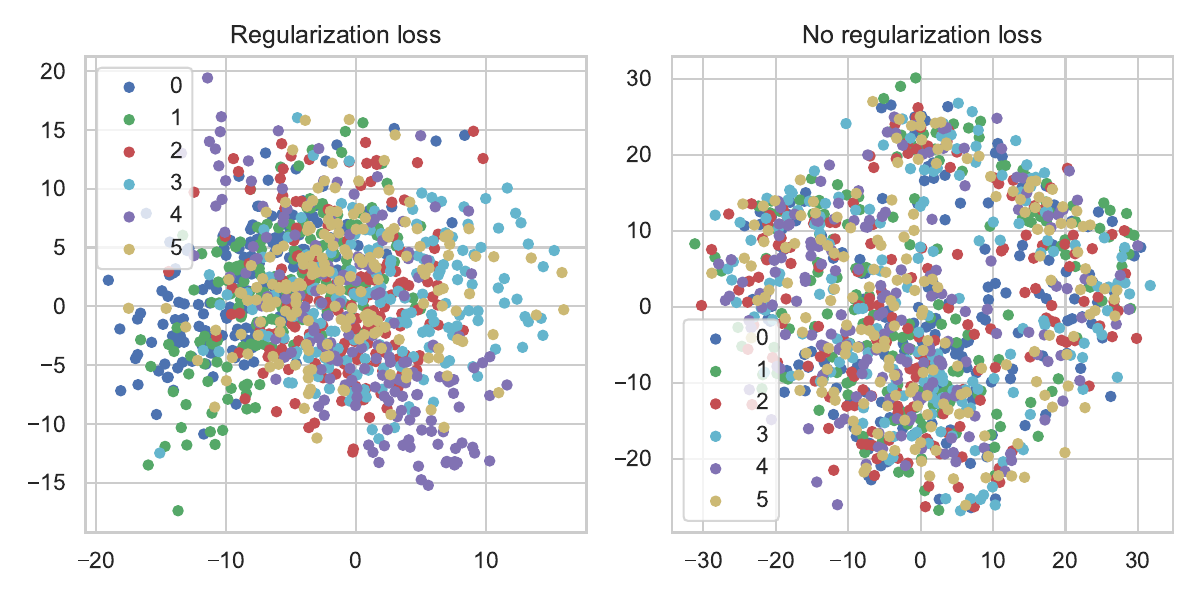}
    \caption{Dimensionality reduction visualization of prompt vectors at each layer of models in the CIKM dataset, where the legend indicates the layer of the pre-trained model.}
    \label{fig:scatter}
\end{figure}

We specifically conduct visual experiments on the effectiveness of the compactness regularization loss using the CIKM dataset. Specifically, we visualize the prompt vectors at different layers of the pre-trained models generated based on personalized information of some users mapped to a two-dimensional space. 

Through experiments, as shown in Figure~\ref{fig:scatter}, we observe significant changes in the distribution of prompt vectors at different layers after incorporating regularization. Specifically, prompts at the same layer tend to merge for different users, while there remains relative diversity across different layers of the pre-trained model. This indicates that the proposed regularization loss can control the diversity of prompt factors and further regulate the diversity of prompt vectors across layers. Consequently, it ensures the progressiveness of our proposed customized prompt learning, where prompts needed at different layers of the model vary, aligning with the assumption of users' interest representation progressively refining across different layers of the model. Additionally, due to user diversity, prompts at the same layer may not be entirely identical for different users, which also aligns with our intuition.

% \subsection{Analysis of Denoising and Attention Block}
% To deeply analyze the relationship between attention blocks and denoising blocks in the pre-trained model, we conduct the following ablation experiments, which include four attention blocks, four denoising blocks, two attention blocks plus two denoising blocks, and the reverse order. The results of the experiment are shown in Figure~\ref{FIG_zhuzhuang}. The findings indicate that the model’s performance is optimized when the denoising blocks are positioned prior to the attention blocks, thus validating the efficacy of our proposed Denoising Pre-training model architecture. Intuitively, the user behavior data is first sufficiently denoised by the filter, and then the complex relationship between the tokens is learned by the attention layer, which leads to the best learning effect of the model.

% \begin{figure*}[h]\centering
% 	\includegraphics[width=0.8\textwidth]{sections/zzt.pdf}
% 	\caption{Ablation study of denoising and attention block.}
%     \label{FIG_zhuzhuang}
% \end{figure*}

\section{CONCLUSION}
In conclusion, our proposed approach, DPCPL, introduced an innovative solution for the efficiency challenges in Multi-Behavior Sequential Recommendation. The Efficient Behavior Miner effectively filtered noise in multi-behavior data, enhancing the model's understanding of contextual semantics. Additionally, the Customized Prompt Learning module enabled highly efficient fine-tuning, leveraging personalized, progressive, and diverse prompts. Extensive experiments on three real-world datasets had unequivocally demonstrated that DPCPL not only exhibited high efficiency and effectiveness, requiring minimal parameter adjustments but also surpassed the state-of-the-art performance across a diverse range of downstream tasks. In the future, we plan to research the integration of finer-grained user interaction data and explore the scalability of efficient denoising modules and customized prompt learning modules in more complex application scenarios.

% % use section* for acknowledgment
% \ifCLASSOPTIONcompsoc
%   % The Computer Society usually uses the plural form
%   \section*{Acknowledgments}
% \else
%   % regular IEEE prefers the singular form
% \fi

\ifCLASSOPTIONcaptionsoff
  \newpage
\fi



\bibliographystyle{IEEEtran}
\bibliography{kdd2024}
\end{document}